\documentclass[5p,12pt, a4paper]{elsarticle}

\usepackage{amssymb}
\usepackage{lineno}

\usepackage{float}
\usepackage{import}

\usepackage{booktabs}

\usepackage[printonlyused]{acronym}
\usepackage{subfiles}
\usepackage{listings}

\usepackage{xcolor}
\definecolor{codegreen}{rgb}{0,0.6,0}
\definecolor{codegray}{rgb}{0.5,0.5,0.5}
\definecolor{codeorange}{rgb}{1,0.49,0}
\definecolor{backcolour}{rgb}{0.95,0.95,0.96}
\lstdefinestyle{mystyle}{
    backgroundcolor=\color{backcolour},
    commentstyle=\color{codegray},
    keywordstyle=\color{codeorange},
    numberstyle=\tiny\color{codegray},
    stringstyle=\color{codegreen},
    basicstyle=\ttfamily\footnotesize,
    breakatwhitespace=false,
    breaklines=true,
    captionpos=b,
    keepspaces=true,
    numbers=left,
    numbersep=5pt,
    showspaces=false,
    showstringspaces=false,
    showtabs=false,
    tabsize=2,
    xleftmargin=10pt,
}

\usepackage{graphicx} % package to use links
\PassOptionsToPackage{hyphens}{url}\usepackage[hidelinks]{hyperref}
\usepackage[nonumberlist,nogroupskip]{glossaries}
\newglossaryentry{clima}{
    name={CBE Clima Tool},
    description={Web based tool to visualize EPW weather files}
}

\restylefloat{table}

\journal{SoftwareX}

\begin{document}

    \begin{frontmatter}
        \title{CBE Clima Tool: a free and open-source web application for climate analysis tailored to sustainable building design}
        \author[label1]{Giovanni Betti\corref{cor1}}
        \author[label2]{Federico Tartarini}
        \author[label3]{Christine Nguyen}
        \author[label1]{Stefano Schiavon}

        \address[label1]{Center for the Built Environment, University of California Berkeley, USA}
        \address[label2]{Berkeley Education Alliance for Research in Singapore, Singapore}
        \address[label3]{College of Letters and Sciences, University of California Berkeley, USA}

        \cortext[cor1]{Corresponding author. E-mail address: gbetti@berkeley.edu}

        \begin{abstract}
            Buildings that are designed specifically to respond to the local climate can be more comfortable, energy-efficient, and with a lower environmental impact. 
            However, there are many social, cultural, and economic obstacles that might prevent the wide adoption of designing climate-adapted buildings. 
            One of the said obstacles can be removed by enabling practitioners to easily access and analyse local climate data.
            The CBE Clima Tool (Clima) is a free and open-source web application that offers easy access to publicly available weather files (in EPW format) specifically created for building energy simulation and design. 
            It provides a series of interactive visualization of the variables therein contained and several derived ones. 
            It is aimed at students, educators, and practitioners in the architecture and engineering fields. 
            Since its launch has been consistently recording over 3000 monthly unique users from over 70 countries worldwide, both in professional and educational settings
        \end{abstract}


    \end{frontmatter}

    \section*{Code Metadata}
    \noindent Current code version: v0.5.0 \\
    Link to repository: \url{https://github.com/CenterForTheBuiltEnvironment/clima} \\
    Legal Code License: MIT \\             
    Code versioning system used: git \\                                        
    Software code language: Python \\
    Documentation: \url{https://center-for-the-built-environment.gitbook.io/clima/} \\
    Support email: \url{https://github.com/CenterForTheBuiltEnvironment/clima/issues}\\
%     \linenumbers

    \section*{Nomenclature}
    % \section*{Nomenclature}
\renewcommand{\baselinestretch}{0.75}\normalsize
\renewcommand{\aclabelfont}[1]{\textsc{\acsfont{#1}}}
\begin{acronym}[longest]

    \acro{t-db}[$T_{db}$]{dry-bulb air temperature\acroextra{, $^{\circ}$C}}
    \acro{t-wb}[$T_{wb}$]{wet-bulb air temperature\acroextra{, $^{\circ}$C}}
    \acro{t-dp}[$T_{dp}$]{dew point air temperature\acroextra{, $^{\circ}$C}}
    \acro{ghi}[$GHI$]{Global Horizontal Irradiance\acroextra{, Wh/m$^2$}}
    \acro{ws}[$U$]{wind speed\acroextra{, m/s}}
    \acro{hr}[$HR$]{Humidity Ratio\acroextra{, kg$_{water}$/kg$_{dry\;air}$}}
    \acro{t-op}[$T_{o}$]{operative air temperature\acroextra{, $^{\circ}$C}}
    \acro{tg}[$t_{g}$]{globe temperature\acroextra{, $^{\circ}$C}}
    \acro{rh}[$RH$]{relative humidity\acroextra{, \%}}
    \acro{v}[$V$]{average air speed\acroextra{, m/s}}
    \acro{t-r}[$\overline{t_{r}}$]{mean radiant temperature\acroextra{, $^{\circ}$C}}
    \acro{clo}[$I_{cl}$]{total clothing insulation\acroextra{, clo}}
    \acro{i-cl}[$i_{cl}$]{permeation efficiency of water vapor through the clothing layer}
    \acro{met}[$M$]{rate of metabolic heat production\acroextra{, W/m\textsuperscript{2}}}

    \acro{e}[$\varepsilon$]{average emissivity of clothing or body surface}
    \acro{sigma}[$\sigma$]{Stefan-Boltzmann constant\acroextra{, 5.67 x 10\textsuperscript{-8} W/(m\textsuperscript{2}K\textsuperscript{2})}}

    \acro{pmv}[PMV]{Predicted Mean Vote}
    \acro{ppd}[PPD]{Predicted Percentage of Dissatisfied\acroextra{, \%}}
    \acro{set}[SET]{Standard Effective Temperature\acroextra{, $^{\circ}$C}}
    \acro{ce}[CE]{Cooling Effect\acroextra{, $^{\circ}$C}}
    \acro{phs}[PHS]{Predicted Heat Strain}

    \acro{BMS}[BMS]{Building Management System}
    \acro{HVAC}[HVAC]{Heating, Ventilation, and Air Conditioning}
    \acro{VAV}[VAV]{Variable Air Volume}
    \acro{AHU}[AHU]{Air Handling Unit}

    \acro{ema}[EMA]{Ecological Momentary Assessment}
    \acro{sdk}[SDK]{Software Development Kit}
    \acro{api}[API]{Application Programming Interface}

    \acro{wmo}[WMO]{World Meteorological Organization}
    \acro{who}[WHO]{World Health Organization}
    \acro{cdc}[CDC]{Centers for Disease Control and Prevention}
    \acro{noaa}[NOAA]{National Oceanic and Atmospheric Administration}
    \acro{epa}[EPA]{United States Environmental Protection Agency}
    \acro{iea}[IEA]{International Energy Agency}
    \acro{un}[UN]{United Nations}
    \acro{epw}[EPW]{EnergyPlus Weather}
    \acro{utci}[UTCI]{Universal Thermal Climate Index\acroextra{, $^{\circ}$C}}
    \acro{svg}[SVG]{Scalable Vector Graphics}
    \acro{csv}[CSV]{Comma-Separated Values}

\end{acronym}
\renewcommand{\baselinestretch}{1}\normalsize

    \section{Motivation and significance}\label{sec:motivation-and-significance}
Buildings construction and operation account for $\approx$~36\% of global final energy use and nearly 40\% of energy-related carbon dioxide (CO\textsubscript{2}) emissions~\cite{InternationalEnergyAgency2018}.
However, a sixth of the occupants of the building are not satisfied with their indoor environment~\cite{Graham2021}.
There is a need to reduce the greenhouse gas emissions related to buildings and increase their resilience while improving occupant well-being.
Energy consumption can be reduced and occupant comfort can be improved when a building is climatically adapted to its environment~\cite{day_gunderson_2015, brager_baker_2009, altomonte_schiavon_kent_brager_2017}. 
The climatically adapted design of a building must take into account the outdoor and indoor climates, optimize site selection, building orientation relative to the sun and the wind, daylight, and view access~\cite{SANTAMOURIS201374, RATTI200349}.
This process is present in vernacular design traditions all over the world~\cite{mileto2014vernacular}.
The early modern movement in architecture took a renewed interest in questions related to solar exposure and orientation, as can be clearly recognized in many sketches, built works, and theoretical writings by Le Corbusier and Alvar Aalto~\cite{menin2003nature}.
After World War II, data and physics-based approaches to buildings climate adaptation started to emerge.
In the early 1950s, the Form and Climate Research Group at the Columbia Graduate School of Architecture sought to develop techniques for refining the design process according to climatic adaptability~\cite{Barber2020MAaC}.
Almost contemporary, the American Institute of Architects started publishing a seminal set of climate graphs to describe the climate of different American cities~\cite{Taylor1949}.
This work about climate data, climatic conditions, and their implication for architectural design was then brought forward by the research of Victor and Aladar Olgyay, among whose many contributions there is the Bioclimatic Chart~\cite{Olgyay1963}.
Shortly after, the emergence of Computer-Aided Design generated a few early tools that aimed at creating climatic data visualizations for architects and building engineers, such as the work by Murray Milne at UCLA~\cite{Milne}.
From 1978 onward, Milne has developed several tools that use climatic data as design guidance, most notably Climate Consultant~\cite{climateconsultant}.
One of the early challenges for the Architecture Engineering and Construction (AEC) industry was the lack of a consistent climate data format. 
This issue was solved with the introduction of the \ac{epw} format~\cite{Drury}.
In the early 2000s, Andrew Marsh developed Ecotect and the Weather Tool, now retired, which have enabled thousands of practitioners and students to integrate climate data in their CAD workflows for the first time~\cite{ECOTECT}.
The Weather Tool shared with Climate Consultant the fact of being a stand-alone, easy-to-use application that enabled the visualization of weather data with some degree of customization.
Recently, new tools have emerged in the field of climate data visualization for the AEC industry, most notably Ladybug Tools~\cite{Ladybug}, Climate Studio~\cite{ClimateS53:online}, and cove.tool~\cite{covetool38:online}.
All of these projects enable different climate visualization directly in the most prevalent CAD environments, mainly Autodesk Revit and Rhinoceros 3D~\cite{mcneel-rhinoceros}.
To achieve this integration, the tools act as software plug-ins.
This strategy has certain advantages, i.e., showing contextual information in the user’s design software of choice.
On the other hand, it makes the tools part of a closed-source ecosystem and creates barriers to their use as they require multiple installations and increasing levels of specialization.
For example, to use Ladybug one needs to be familiar with Grasshopper, the visual programming extension to Rhinoceros 3D~\cite{mcneel-rhinoceros}. 
Existing stand-alone tools, like Climate Consultant, do not provide vector graphics and therefore their use in practice and education is limited~\cite{climateanalysistool}. 

Our aim is to develop open-source software that overcomes the limitations listed above, reduce barriers to access, and helps users to design and construct healthier and sustainable buildings.

\section{Software description}\label{sec:software-description}
The \gls{clima} is a free open-source web application for the analysis and visualization of climate data specifically designed to support the needs of architects, engineers, students, and educators interested in climate-adapted building design.
It does not only visualize data, but it organizes, manipulates, and relates them in an easy-to-interpret fashion.
The landing page is characterized by an interactive world map.
Through this map, the user can access circa 30,000 publicly available weather data files.
Data are collected from publicly available repositories maintained by EnergyPlus and Climate.OneBuilding.org.
Alternatively, the user can upload any valid \ac{epw} file. 
The uploaded file is automatically processed and analyzed.
Our tool does not perform quality checks other than checking that the data supplied matches the expected structure of an \ac{epw} file. 
In case of a mismatch, an error message is shown to the user.

\begin{figure}[htb!]
    \centering
    \includegraphics[width=\columnwidth]{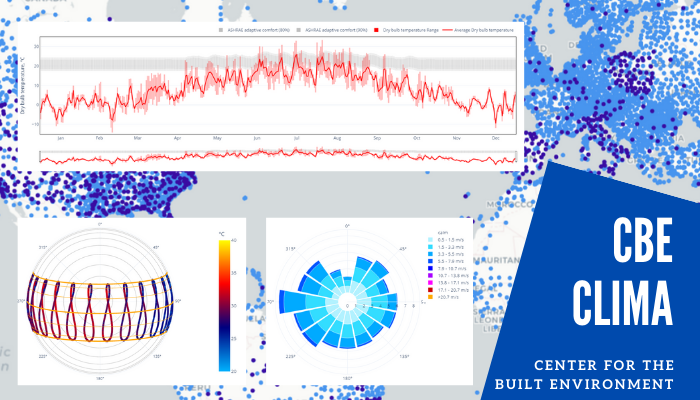}
    \caption{Video tutorial -- An introduction to the CBE Clima Tool.}
    \label{fig:video}
\end{figure}

The data contained within the \ac{epw} files are used to generate easy-to-interpret charts.
Many of these allow the user to explore data at different temporal resolutions, highlighting seasonal, daily, or hourly patterns.
All the charts allow some degree of interactivity, allowing the user to hover, zoom, and highlight data.
For some charts, we also allow the user to select specific time ranges (i.e. wind roses analysis).
Not all relevant data for a climate analysis is contained directly in the \ac{epw} files but can be derived from the information contained therein.
We use the open-source Python library \verb|pvlib|~\cite{F.Holmgren2018} to calculate the apparent solar position (altitude and azimuth).
We use the open-source Python library \verb|pythermalcomfort|~\cite{pythermalcomfort} to calculate other indices that can be derived from those contained in the EPW files.
For example, we calculate the \ac{utci}~\cite{utci} scenarios to understand outdoor comfort, solar gain on occupants~\cite{Arens2015}, the adaptive thermal comfort model~\cite{ASHRAE552020}, natural ventilation potential, heating and cooling degree days, and various other psychrometric variables.
The \gls{clima} can be accessed at \url{https://clima.cbe.berkeley.edu}.

\subsection{Software Architecture}\label{subsec:software-architecture}
\begin{figure*}[htb!]
    \centering
    \includegraphics[width=\textwidth]{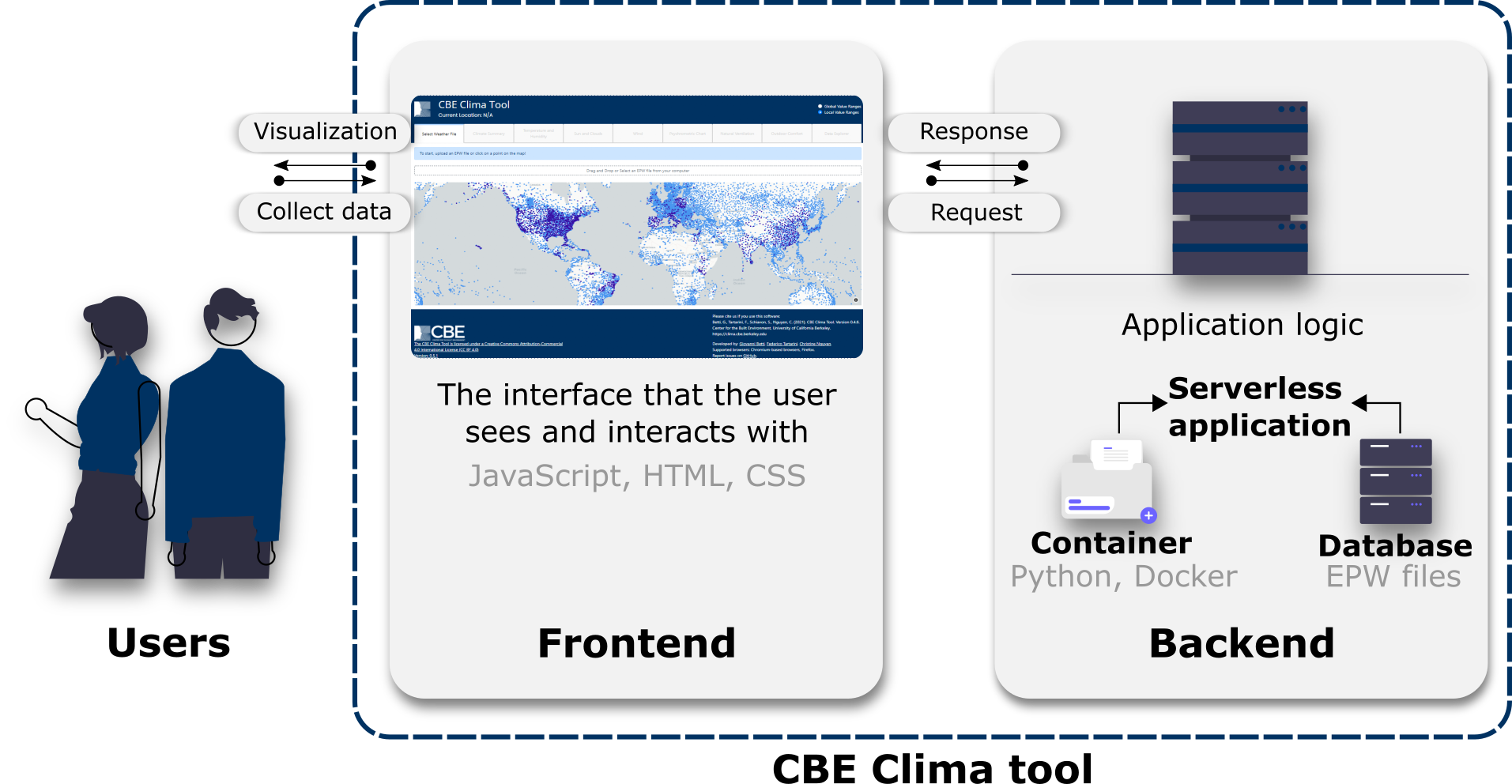}
    \caption{Application architecture diagram.}
    \label{fig:application_arch}
\end{figure*}

The \gls{clima} tool is built using Dash~\cite{plotly} a Python framework to build web analytic applications and data visualization websites.
We use Dash since it is an open-source library released under an MIT license, and applications built with it can be rendered in any web browser.
Clima is, therefore, cross-platform and compatible with the most widely used web browsers.
Dash applications are web servers that run Flask and communicate JSON packets over HTTP requests.
We have deployed \gls{clima} using a cloud service that hosts our containerized application.
We use Docker to create a container that listens to user requests.
A request is triggered each time the user navigates to the URL: \url{https://clima.cbe.berkeley.edu}.
Dash’s front-end renders components using React.js, a Javascript framework.
While the back-end code runs on a cloud instance, the front-end code is run locally inside the user's browser.
Each time a user changes some of the input parameters, a new request is sent to the server.
The Python source code is stored inside the container, while the EPW data is queried from the databases maintained by EnergyPlus and Climate.OneBuilding.org.
A diagram depicting the application architecture is shown in Figure~\ref{fig:application_arch}.
For more information about how a Dash application works please visit and consult their official documentation.

\subsection{Software Functionalities}\label{subsec:software-functionalities}
\begin{figure*}[htb!]
    \centering
    \includegraphics[width=\textwidth]{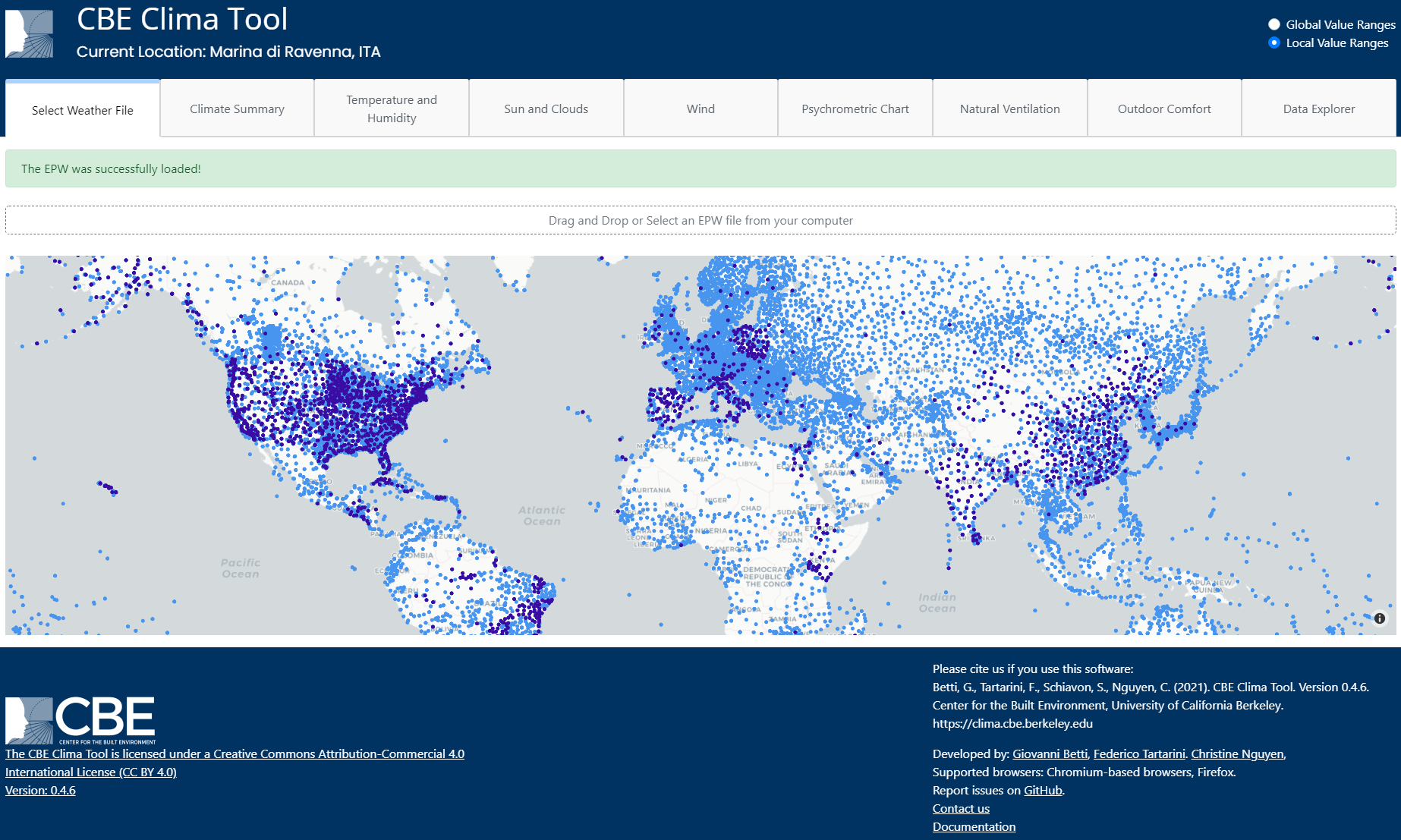}
    \caption{\gls{clima} home page where the \ac{epw} file is selected.}
    \label{fig:clima_home}
\end{figure*}

The \gls{clima} allows users to analyze and visualize climatic data.
Users can access \ac{epw} files via a map-based interface (as previously stated in Section~\ref{sec:software-description}).
All images and charts included in our tool can be downloaded in \ac{svg} format.
This allows them to be modified and adapted according to the users' needs.
Furthermore, the vector nature of the \ac{svg} format enables both high-resolution on-screen display and high-quality printing.
This is an important feature of the tool as it enables visually compelling data communication with multiple stakeholders.
The interactive nature of the graphs means that the user can inspect the underlying numerical information through rich mouse-over events.
Users can download, in \ac{csv} format, either the \ac{epw} file or the \gls{clima} data frame created in the app back-end to generate the plots.
The latter contains the variables in the \ac{epw} and additional ones derived from it, as previously described.
When displaying the variables in the interactive plots, users can either select to use the Global or Local Value Ranges, using the radio button at the top of the page, see Figure~\ref{fig:clima_home}.
The `Global' option uses preset limits for the chart axes, chosen to cover the vast majority of the climatic ranges to be found on Earth.
This allows users to effectively compare charts generated for different locations.
The `Local' option sets the upper and lower limits of the chart axes as a function of the data contained in the \ac{epw} file.

The following paragraphs give an overview of the main functionalities of each page of the tool.
However, we recommend consulting the official project documentation for more detailed and up-to-date information available at \url{https://cbe-berkeley.gitbook.io/clima/}.

\paragraph{Home Page} Users can either choose to analyze the climate of the locations displayed on the map or upload a custom \ac{epw} file as shown in Figure~\ref{fig:clima_home}.
After loading an \ac{epw} file the user can then access the other tabs.

\paragraph{Climate Summary} It provides textual information about the selected location such as longitude, latitude, elevation, Koppen-Geiger climate zone, average/hottest/coldest yearly temperatures, annual cumulative horizontal solar radiation, and percentage of diffuse horizontal solar radiation.
The user is presented with a further map showing the location of the weather station selected.
The user can also download the source data, visualize the heating and cooling degree day chart and compute them as a function of user-defined base temperatures and analyze the distributions of the \ac{t-db}, \ac{rh}, \ac{ghi}, and \ac{ws}.

\paragraph{Temperature and Humidity} It shows the yearly, hourly, and heat-map charts for \ac{t-db} and \ac{rh}.
The yearly \ac{t-db} chart depicts the daily temperature range and the mean value together with the ASHRAE adaptive thermal comfort region for the 80\% and 90\% acceptability ranges.
The prevailing mean outdoor temperature was calculated using an exponentially weighted running mean of the daily outdoor temperature for the last 7 days and assuming an $\alpha$ of 0.9~\cite{EN1679812019}.
The daily chart shows how the hourly readings vary as a function of the hour of the day, and data are grouped by month.
The heat map shows each individual hourly data point in a grid where the vertical axis shows the time of the day, while the horizontal axis shows the day of the year. 
The table with descriptive statistics for each month and year is shown at the bottom of the page.

\paragraph{Sun and Clouds} It visualizes the apparent sun path as seen by an observer at the weather station location either in Spherical coordinates (where the observer sits at the weather station location and the apparent altitude and azimuth are mapped to the polar and azimuthal angles respectively) or Cartesian coordinates (where the apparent altitude and azimuth are mapped to a Cartesian plot).
The monthly average hourly global and diffuse solar radiation (kWh/m\textsuperscript{2}), and the monthly cloud coverage are also presented.
Users can also plot and visualize how any of the many variables related to solar radiation contained in the \ac{epw} file (e.g. \ac{ghi}, extraterrestrial horizontal irradiation) vary throughout the year.

\paragraph{Wind} It provides information about the wind velocity and direction for the selected location.
It comprises several wind roses that either display the full dataset (yearly data), values grouped by seasons or grouped by time of day.
The user can also create a custom wind rose by filtering the data by month and hour of the day.
This page also shows two heat maps, one for the wind speed and one for the wind direction.

\paragraph{Psychrometric Chart} It shows data on a customizable psychrometric chart.
Users can select to bin the data and colour them by frequency or display the data and colour them by any variable of choice.
Optionally, filters can be applied to select data contained in a specific time range, or filtered according to a secondary variable.
As an example, see the figure in Section~\ref{subsec:psy} for more.

\paragraph{Natural Ventilation} It allows the user to estimate when and for how long the climate is suitable for the use of natural ventilation.
This can be achieved by selecting an outdoor \ac{t-db} range in which natural ventilation can be used.
The user can also filter data either based on monthly or hourly ranges.
If radiant systems are used to cool the building, the user can also specify the surface temperature at which they operate.
The \gls{clima} filters out automatically data when the surface of the radiant system is below the dew point.

\paragraph{Outdoor comfort} It plots the \ac{utci} values on two heat-maps.
The first chart shows the \ac{utci} equivalent temperature, while the second shows the data binned into the 10 \ac{utci} stress categories (from extreme cold stress to extreme hot stress).
Four distinct \ac{utci} scenarios are calculated showing the impact of exposure to the sun and the wind and combinations thereof.
This might help the designer to understand what combination of exposure to or protection from the sun and wind would provide the most comfortable outdoor environment at different times of the year.

\begin{figure*}[htb!]
    \centering
    \includegraphics[width=\textwidth]{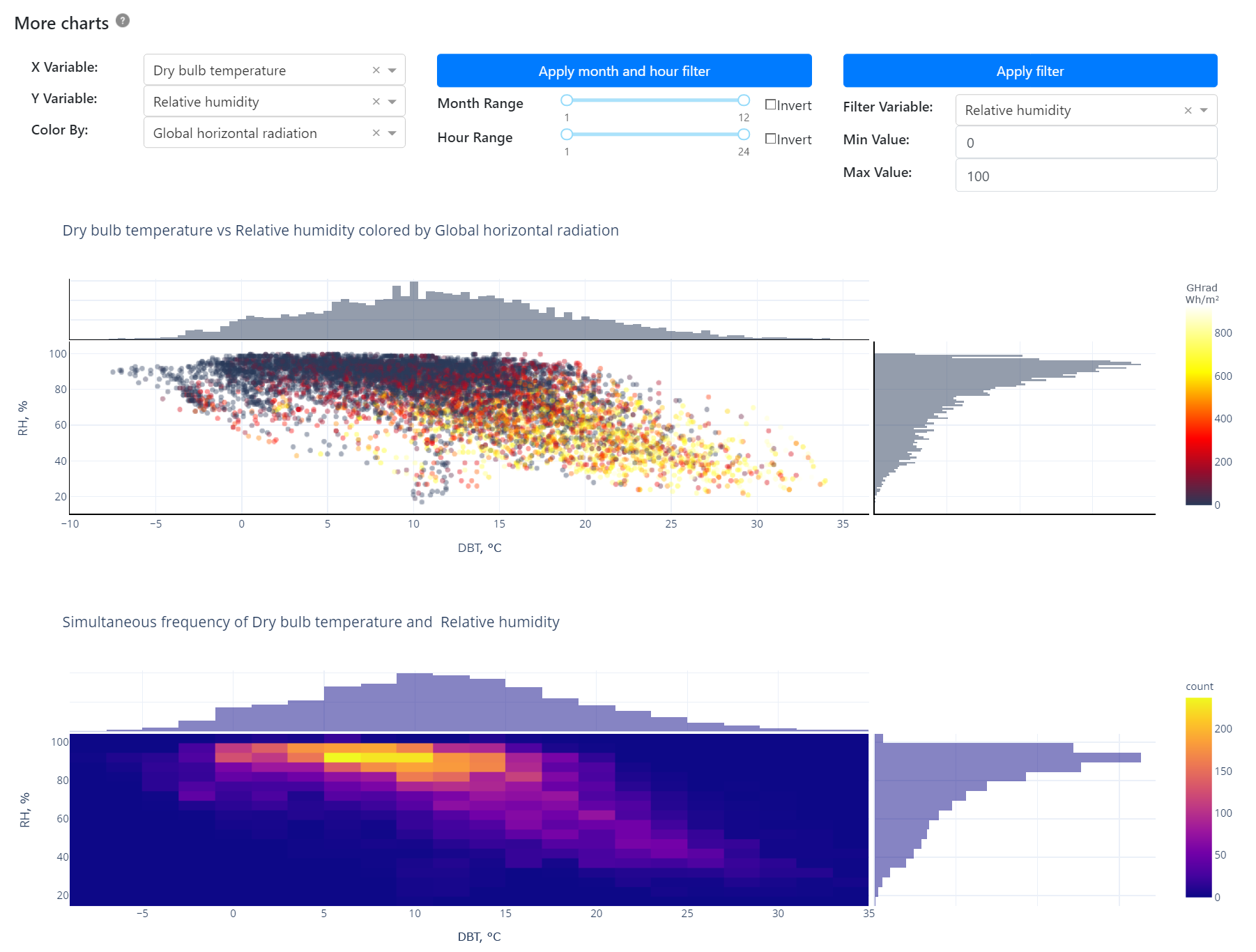}
    \caption{Data explorer tab -- The user can select two variables one to be displayed on the x-axis and one to be displayed on the y-axis.
    This allows the user to visualize the correlation between the two variables.
    While the histograms depict the distribution of the data.
    This selection is then used to create a scatter plot and a heat-map.
    In the scatter plot a third variable is used to color the markers and the range is displayed in the color bar.
    The heat-map, on the other hand, depicts the correlation between the x and y variables and shows the number of points that are comprised in each bin.
    }
    \label{fig:clima_explore}
\end{figure*}

\paragraph{Data Explorer} It allows the user to explore and plot all variables contained in the source dataset.
It comprises three charts (yearly chart, daily chart, and heat-map chart) that are also displayed on the `Temperature and Humidity' page.
Moreover, it also allows users to visualize the correlation between three input variables.
For example in Figure~\ref{fig:clima_explore} is a screenshot of the aforementioned charts, and we are showing how data points are distributed as a function of \ac{t-db}, \ac{rh}, and \ac{ghi}.

\section{Illustrative examples}\label{sec:illustrative-examples}

\subsection{Sun Path Chart}
\begin{figure}[htb!]
    \centering
    \includegraphics[width=\columnwidth]{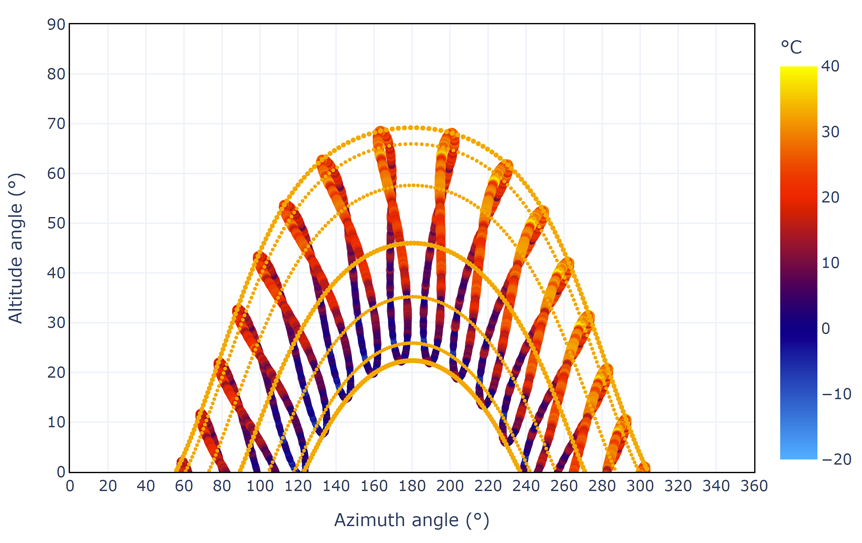}
    \caption{Cartesian sun path chart.
    It can be used to determine the solar azimuth and altitude for a given location.
    In addition, we allow users to colour the points using a third variable.
    This allows for example to determine how the \ac{t-db} varies as a function of the solar azimuth and altitude throughout the year.}
    \label{fig:clima_sun_path}
\end{figure}

Sun control and shading systems can significantly reduce peak heat gain and cooling requirements, while also improving the quality of natural lighting in the interior of the building.
This example shows how the user can visualize how \ac{t-db} varies as a function of the position of the sun in the sky.
This information can then be used to optimize the design of sun-shading devices.
This chart can be generated from the `Sun and Clouds` tab.
The user first has to select, using the drop-down, between a spherical or Cartesian coordinate system.
In this example, we use the latter and decided to select \ac{t-db} as a secondary variable to plot from the second dropdown.
Figure~\ref{fig:clima_sun_path} shows that this location is characterized by cold winters and hot summers.
The user can then determine the optimal size and shape of the shading that can be used to minimize solar gains in summer while maximizing passive solar heating in winter.

\subsection{Psychrometric Chart}\label{subsec:psy}
\begin{figure}[htb!]
    \centering
    \includegraphics[width=\columnwidth]{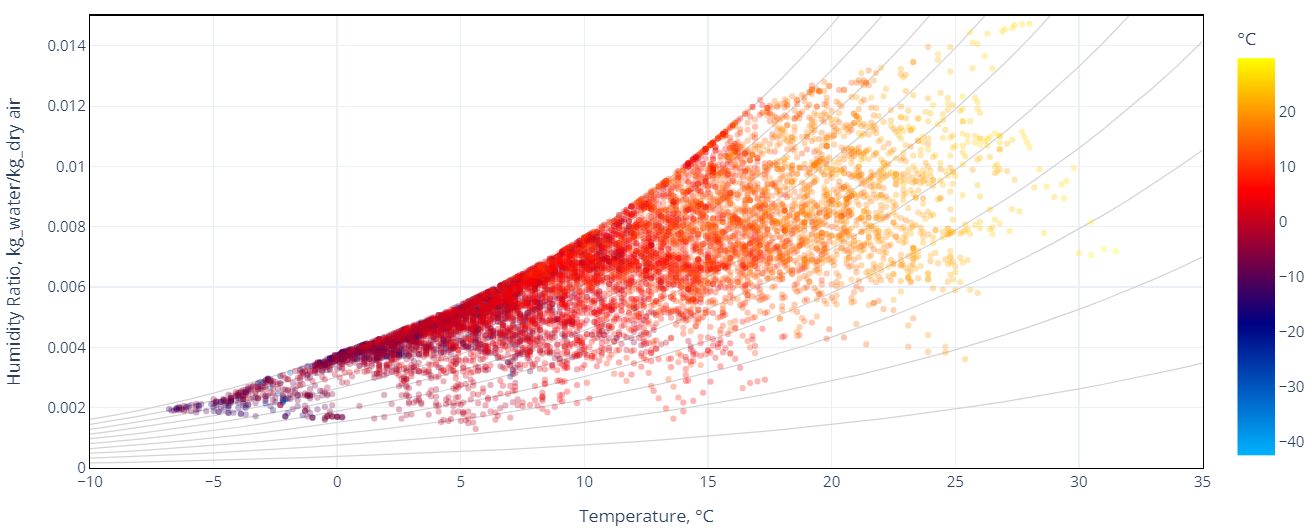}
    \caption{Psychrometric chart.
    It can be used to visualize the relationship between dry air and moisture.
    In addition, we allow users to colour the points using a third variable, in this example, we are showing how the \ac{utci} -- No Sun and Wind -- varies as a function of the dry-bulb air temperature and the humidity ratio.}
    \label{fig:clima_psy}
\end{figure}

This example shows how the user can visualize the distribution of the climate data for a specific location on the psychrometric chart.
We also decided to colour the data points as a function of the \ac{utci} calculated for the condition where the reference person is neither exposed to direct solar radiation nor to wind.
This information can be used to design a semi-outdoor space or to determine when natural ventilation may be used.
This chart can be generated from the `Psychrometric` tab.
Through the drop-down located on the top left side of the page, the user can select the variable `UTCI: no Sun \& no Wind`.
The \gls{clima} automatically generates the chart shown in Figure~\ref{fig:clima_psy}.

\section{Impact}\label{sec:impact}
The \gls{clima} is intended to be used by architects, engineers, researchers, educators, and students.
It can even be used by non-technical users with no analytical or programming skills.
Compared with similar competing solutions for climate analysis, \gls{clima} offers a very complete and flexible feature set that makes it useful for a varied group of users.
The \gls{clima} can be used by students exposed to this topic for the first time and by designers with many years of experience.
The easy output of high-quality vector graphics licensed under Creative Commons Attribution 4.0 International (CC BY 4.0) facilitates its integration into presentation documents.  
While alternative solutions require either subscriptions or one or more installs, \gls{clima} can be used via any device equipped with a browser.
Some of the most unique features of the tool, in addition to the completeness, interactivity and customizability of the data visualizations produced, pertain to the availability of complex metrics such as Natural Ventilation potential and the Universal Thermal Climate Index (UTCI) under four distinct scenarios.  
We have also open-sourced the code so users around the world can contribute to our project or use our functions in their applications.
The source code is licensed under the MIT license.
This permissive license enables users to share, copy, and redistribute the material in any medium or format and to adapt, remix, transform, and build upon the material for any purpose, even commercially.
These choices were dictated in the hope of promoting the widest possible adoption.
We released the first version of this tool in August 2021.
Since its release, the tool has consistently gathered over 2000 organic unique monthly users from more than 70 nations worldwide.

\section{Conclusions}\label{sec:conclusions}
The \gls{clima} is a free and open-source web application for the analysis and visualization of climate data specifically designed to support the needs of architects, engineers, students, and educators interested in the design of climate-adapted buildings. 
It aims to reduce barriers to the implementation of sustainable design.
\gls{clima} processes and visualizes \ac{epw} climate files. 
It allows users to perform complex psychrometric data calculations and interactively visualize the results without the need of writing any code or installing specialized software. 
It is written in Python. 
All generated charts can be downloaded in \ac{svg} format; this allows them to be modified and adapted according to the users' needs.
It is composed of the following main pages: home, climate summary, temperature and humidity, sun and clouds, psychrometric chart, natural ventilation, outdoor comfort, and data explorer. The pages contain interactive graphs and tables. 
The tool allows an easy and accessible interpretation of climate data. 

\section{Conflict of Interest}\label{sec:conflict-of-interest}
We confirm that there are no known conflicts of interest associated with this publication and that there has not been financial support for this work that could have influenced its outcome.

\section*{Acknowledgements}
We would like to acknowledge the work of the authors who contributed to the development of the \gls{clima}\footnote{\url{https://github.com/CenterForTheBuiltEnvironment/clima/graphs/contributors}}.
This research has been supported by the Center for the Built Environment at the University of California Berkeley and the Republic of Singapore’s National Research Foundation through a grant to the Berkeley Education Alliance for Research in Singapore (BEARS) for the Singapore-Berkeley Building Efficiency and Sustainability in the Tropics (SinBerBEST) Program.

\bibliographystyle{elsarticle-num}
\bibliography{ms}

\end{document}